\documentclass[preprint,showpacs,preprintnumbers,amsmath,amssymb, showkeys]{revtex4}

\usepackage{array}
\usepackage{booktabs}
\usepackage{tabu}
\usepackage{dcolumn}
\usepackage{amsmath}
\usepackage{amsfonts}
\usepackage{amssymb}
\usepackage{graphicx}
\usepackage{subfigure}
\usepackage{graphicx}
\usepackage{dcolumn}
\usepackage{bm}
\RequirePackage[colorlinks,citecolor=red,urlcolor=red,linkcolor=blue]{hyperref}

\begin{document}

\title{Constant-roll approach to non-canonical inflation}

\author{Abolhassan Mohammadi$^{1,2}$}
  \email{abolhassanm@gmail.com}
\author{Khaled Saaidi$^1$}
  \email{ksaaidi@uok.ac.ir}
  \author{Haidar Sheikhahmadi$^{3,4}$}
  \email{h.sh.ahmadi@gmail.com}

\affiliation{$^1$Department of Physics, University of Kurdistan, Pasdaran Street, P.O. Box 66177-15175, Sanandaj, Iran\\
$^2$Dipartimento di Fisica e Astronomia, Alma Mater Universit\`{a} di Bologna, via Irnerio 46, 40126 Bologna, Italy\\
$^3$Center for Space Research, North-West University, Mafikeng, South Africa\\
$^4$School of Astronomy, Institute for Research in Fundamental Sciences (IPM), P. O. Box 19395-5531, Tehran, Iran}
\date{\today}

\begin{abstract}
The scenario of constant-roll inflation in the frame work of a non-canonical inflaton model will be studied. Both of these modifications lead to appearance of some differences in the slow-roll parameters besides the Friedmann equations resulted in a better justification of theoretical predictions comparing to the observation. Phenomenologically, by assuming a constant $\eta$, i.e. second slow roll parameter,  and recalculating the related perturbation equations obviously there should appear some modification  in the scalar spectral index and amplitude of scalar perturbations. It will be shown that finding an exact solution for Hubble parameter is one of the main advantages and triumphs of this approach. Also, whereas making a connection between sub-horizon and super-horizon regions has a crucial role in inflationary studies the main perturbation parameters will be obtained at the horizon crossing time. To examine the accuracy of our results we shall consider the Planck 2018 results as a confident criterion. To do so by virtue of the $r-n_s$ diagram, the acceptable ranges of the free parameters of the model will be illustrated.  As a result it will be found out the second slow-roll parameter should be a positive constant and smaller than unity. By constraining the free parameters of the model, also the energy scale of inflation will be estimated that is of order $10^{-2}$. Even more, by investigating the attractor behavior of the model it will be cleared that the aforementioned properties could be appropriately satisfied.
\end{abstract}
\maketitle

\section{Introduction}
The first ideas about the very early {Universe} to cope with (at least) three fundamental problems of the standard Big Bang model dates back to 1981 and the paper proposed by Guth  \cite{Guth},  called "old inflation".  But this model faced a problem to justify the smooth exit of the end of inflation. To solve this fatal problem many models of inflation have been put forward such as new inflation \cite{Linde,Albrecht}, chaotic inflation \cite{Lindea}, k-inflation \cite{Armendariz-Picon,Garriga}, brane inflation \cite{Maartens,Golanbari}, G-inflation \cite{Abolhasani,Maeda,Alexander,Tirandari} warm inflation \cite{Berera,Bereraa,Taylor,Hall,Bastero,Sayar,Akhtari}, etc.  Inflation is assumed to be a phase of a very rapid accelerated expansion of  the super high energy Universe. In single or multiple field scenarios of inflation usually the Universe evolved in the presence of scalar fields \cite{Chen:2009we,Chen:2017ryl}. Nonetheless, there are some models in which instead of scalar field for instance the gauge fields play the main role \cite{Golovnev:2008cf,Adshead:2012kp}. \\
Confidently, the inflationary models almost are based on the slow-roll assumption, in which the scalar field undergoes and rolls down slowly from the top of its potential to the bottom of the hill. The flatness of the potential provides a condition for having a quasi-de Sitter expansion, weak dependency on time of Hubble parameter, especially at the beginning of the inflation. Besides the aforementioned necessary parts to run inflation we need the enough smallness of slow-roll parameters to cope with the well known three problems of the hot Big Bang theory \cite{Lindeb}. The definition of the first slow-roll parameter is  based on the time evolution of Hubble parameter divided by its square, i.e. $\epsilon=-{\dot{H} / H^2}$, and the condition $\epsilon < 1$ is required to have an accelerated expansion phase at the initiation eras ($\ddot{a}>0$). The second slow-roll parameter is  defined as $\eta = {\ddot{\phi} / H\dot{\phi}}$  in which holds out the rate of the time derivative of scalar field during a Hubble time \cite{weinberg}. Smallness of the latter parameter states that $\dot\phi$ should vary very slowly, and therefore guarantees an appropriate behaviour of inflation. \\
Whereas the single field canonical versions of inflation could not  able to cover all results risen by observations one might seek a modification in inflationary models \cite{PhysRevLett112011302,Emami:2013lma,Sheikhahmadi:2019xkx}. One possible extension for instance is quasi-single field inflation or maybe multiple field inflation\cite{Baumann:2011nk,Chen:2012ge,Sefusatti:2012ye}, etc.
Another proposal of scalar field where the Lagrangian is a function of the scalar field $\phi$ and the kinetic term $X={1 \over 2} g^{\mu\nu} \partial_\mu \phi \partial_\nu \phi $; i.e. $\mathcal{L}(\phi,X)$ is the k-essence model \cite{Armendariz-Picon,Garriga,Baumann}.  In this model, the   sound speed is no longer equal to the  light speed, and it could be smaller which leads to a smaller tensor-to-scalar ratio to meet the observational range. One type of aforementioned Lagrngian are those with modified kinetic term of the canonical scalar field Lagrangian. In the present work, the modification to the kinetic term is restricted to the time derivative of the scalar field so that it is possible to receive the canonical one. This type of the Lagrangian has received attention to be applied for cosmological studies, for instance one can see \cite{1a}. In this reference,  the authors studied the dynamical system of the Universe when it is filled with a barotropic fluid and a scalar field with modified kinetic term. Also, in \cite{2a},  a non-canonical scalar field with a modified Lagrangian has bee taken as a candidate for dark energy. It has claimed that for a simple choice of the modified kinetic term, this model can be considered as a unified model of dark matter and dark energy. Besides, these models which usually are based on modification in kinetic portion  have received more attention recently cause of their ability to justify the inflation behaviour. In \cite{3a}, the authors have shown that for specific choices of the free parameters the tensor-to-scalar ratio could increase which originated from enhancement of the sound speed. It has indicated that other consequences of this model can be enumerates as the larger energy scales and higher temperatures for the reheating \cite{5a}. In the latter reference one can find some good clues about the intermediate inflation for non-canonical model with power-law kinetic term. Additionally the authors in \cite{6a} have approved that despite the canonical intermediate inflation, non-canonical model could properly satisfy the observational data in their case for chaotic inflation. The quantities of the model are derived, and the existence of an inflationary attractor is confirmed. Even more, the authors have shown that  for a steep potential, the model is able to properly describe inflation. In addition, results of \cite{7a} have clarified that the power-law inflation in standard model of inflation leads to a tensor-to-scalar ratio which is out of range compared to observational data. However, by using a non-canonical scalar field there could be a new power-law inflation which its predictions are in consistency with Planck results. In \cite{8a} the scenario of warm inflation is extended to the non-canonical case, where it means the kinetic term will be modified again. The new, but still scale invariant, curvature spectrum is obtained and it is demonstrated that the tensor-to-scalar ratio is insignificant for strong regime and significant for weak regime. \\ 
Such models have a better consistency with observational data compared to canonical scalar field models. Some interesting features of non-canonical scalar field model can be addressed as follows \cite{Unnikrishnan}
\begin{itemize}
  \item Steep potentials like $v(\phi) \propto \phi^{-n}$, which are known as the dark energy potential, could give a better inflaton potential in non-canonical scalar field compared to the canonical ones.
  \item The consistency relation $r = -8n_T$ is violated in non-canonical scalar field model of inflation.
  \item In the canonical scalar field models of inflation, the exponential potential roughly stands in an acceptable range of data. However, this type of potential in the non-canonical model of inflation could show a better agreement with data.
\end{itemize}
The slow-roll features of inflation almost is provided by a potential with a flat part. It rises to this question what happens if the potential is exactly flat?. This question for the first time was considered in \cite{Kinney}. From the equation of motion of scalar field, it is determined that for a flat potential the second slow-roll parameters becomes $\eta=-3$, which is not smaller than unity, actually it is of order one. After that, in \cite{Namjoo} the authors have studied the non-Guassianity of the case and it was specified that the non-Gaussianity is not ignorable anymore and it could be of order one \cite{Namjoo}. The idea of having flat potential was generalized in \cite{Martin}, where the authors assumed $\eta$ could be a constant. They found an approximate solution for the model and obtained a scalar perturbation amplitude that could even varies on superhorizon scales, and also for some choices it could be scale invariant. In \cite{Motohashi}, where for the first time the name "constant-roll" was addressed, the same model was reconsider by using Hamilton-Jacobi formalism \cite{Salopek,Liddle,Kinneya,Guo,Aghamohammadi,Saaidi,Sheikhahmadi} and they found an exact solution for Hubble parameter which possesses the attractor behavior as well. Also, it was concluded that the power spectrum could remain scale invariant for specific choices of the constant. The scenario of constant-roll inflation in modified gravity was studied in \cite{Motohashi-b,Nojiri,Odintsov-c,Oikonomou-a,karam}, and the generalized version of this approach, known as smooth-roll inflation could be found in \cite{Odintsov-a,Odintsov-b,Oikonomou-b}. \\
The interesting feature and application of non-canonical scalar field in slow-roll inflationary scenarios motivated us to investigate this model for constant-roll inflation. The perturbation equations will be reconsidered  and we will find the modified amplitude of scalar perturbations, and also the correction to the scalar spectral index are determined which are of the second order of $\eta$. Comparing the result with observational data showed that for some specific values of $\eta$ one could have obtain a scale invariant perturbation on super horizon scale. \\
The paper is organized as follow: In Sec.II, the general formulae of the frame work of non-canonical scalar field will be obtained and they are rewritten for specific choices of the Kinetic term. The slow-roll parameters and the differential equation for the Hubble parameter will be addressed in Sec.III, where the constant-roll approach is applied on the equations. The scalar and tensor perturbations will be discussed in Sec.IV, and the power spectrum of perturbations are derived. And finally We conclude and discuss our results in Section in Sec.V.\\

\section{Non-canonical scalar field model}
The action is assumed to be
\begin{equation}\label{action}
S = {-1 \over 16\pi G} \; \int d^4x \sqrt{-g} \; R + \int d^4x \sqrt{-g} \mathcal{L}(\phi,X)
\end{equation}
where $X = g^{\mu\nu} \partial_\mu \phi \partial_\nu \phi /2$, and $\mathcal{L}(\phi,X)$ is the Lagrangian of non-canonical scalar field that in general is an arbitrary function of scalar field $\phi$ and $X$. Variation of at action with respect to the metric comes to the field equation of the model
\begin{equation}\label{fieldequation}
 R_{\mu\nu} - {1 \over 2}\; g_{\mu\nu} R = 8\pi G \left( {\partial \mathcal{L} \over \partial X}\; \partial_\mu \phi \partial_\nu \phi - g_{\mu\nu} \mathcal{L}  \right)
\end{equation}
and also variation of the action with respect to the field come to the following equation of motion
\begin{equation}\label{EoM}
{\partial \mathcal{L} \over \partial \phi} -
{1 \over \sqrt{-g}}\; \partial_\mu\left( \sqrt{-g} \; {\partial \mathcal{L} \over \partial \big( \partial_\mu \phi \big)} \right) = 0.
\end{equation}
It is assumed that the geometry of the universe is describe by a spatially flat FLRW metric
\begin{equation}
ds^2 = dt^2-a^{2}(t)\; \left(dx^2 + dy^2 + dz^2 \right)
\end{equation}
Since the term in parenthesis on the right hand of the field equation (\ref{action}) is the energy-momentum tensor of scalar field, in a comparison to the energy-momentum of a perfect fluid $T_{\mu\nu} = (\rho + p) u_\mu u_\nu - p g_{\mu\nu}$ it is concluded that the energy density, pressure and four velocity of the field are
\begin{equation}\label{energypressure}
\rho = 2X {\partial \mathcal{L} \over \partial X} - \mathcal{L}, \qquad p = \mathcal{L}, \qquad u_\mu = {\partial_\mu \phi \over \sqrt{2X}}.
\end{equation}
Substituting this metric into the field equation (\ref{fieldequation}), one arrives at the Friedmann equations
\begin{equation}\label{friedmann}
H^2 = {8\pi G \over 3} \; \rho, \qquad \dot{H} = -4\pi G \Big( \rho + p \Big).
\end{equation}
Also, the equation of motion (\ref{EoM}) for this geometry is read as \cite{Unnikrishnan}
\begin{equation}\label{eoms}
\left( {\partial \mathcal{L} \over \partial X} + 2X {\partial^2 \mathcal{L} \over \partial X^2} \right) \ddot{\phi} +
\left( 3H {\partial \mathcal{L} \over \partial X} + \dot{\phi}\; {\partial \mathcal{L} \over \partial X \partial \phi} \right) \dot{\phi} -
{\partial \mathcal{L} \over \partial \phi} = 0
\end{equation}

In the present work, we are going to work with a specific type of k-essence Lagrangian which contains  a modification in the kinetic term. We notice  this model comes to interesting results working out inflation besides the studies of dynamical system of the Universe and dark energy. The Lagrangian is assumed to be 
\begin{equation}\label{Lagrangian}
  \mathcal{L} = X\; \left( {X \over M^4} \right)^{\alpha-1} - V(\phi),
\end{equation}
where $\alpha$ is a dimensionless constant and $M$ is a constant with mass dimension. Using this definition for scalar field Lagrangian, its energy density and pressure are expressed by
\begin{equation}\label{rhopressure}
\rho = (2\alpha - 1) X \left( {X \over M^4} \right)^{\alpha-1} + V(\phi), \qquad p = X\; \left( {X \over M^4} \right)^{\alpha-1} - V(\phi).
\end{equation}
Then, the equation of motion (\ref{eoms}) is reduced to
\begin{equation}\label{eomphi}
\ddot{\phi} + {3H \over 2\alpha-1} \; \dot\phi + \left( {2 M^4 \over \dot\phi^2} \right)^{\alpha-1} {V_\phi(\phi) \over \alpha(2\alpha-1)} =0.
\end{equation}
Introducing the Hubble parameters as a function of scalar field, and using Eq.(\ref{friedmann}), (\ref{rhopressure}) and relation $\dot{H}=\dot{\phi} H_\phi(\phi)$, the time derivatives of the scalar field is given by
\begin{equation}
\dot\phi^{2\alpha-1} = {-2^\alpha M_p^{2(2\alpha-1)} \mu^{4(\alpha-1)} \over \alpha}\; H_{,\phi}(\phi),
\end{equation}
in which $M_p^2 = 1/8\pi G$ and $\mu \equiv M/M_p$. Then substituting this in the Friedmann equation (\ref{friedmann}), and using Eq.(\ref{rhopressure}) the potential of scalar field is read as
\begin{equation}\label{potential}
V(\phi) = 3M_p^2 H^2(\phi) - {(2\alpha - 1)M_p^{4\alpha} \over 2^\alpha M^{4(\alpha-1)}}
\left[ {-2^\alpha \mu^{4(\alpha-1)} \over \alpha}\; H_{,\phi}(\phi) \right]^{2\alpha \over 2\alpha-1}
\end{equation}
For the rest of the paper, the reduced mass planck is taken as $M_p=1$ for more convenience, and also we defined the new parameter $\lambda = M^{4(\alpha-1)}$.

\section{Non-canonical inflation}
During the inflation, the universe undergoes an extreme expansion in short period of time. Here, it is assumed that the inflation is caused by a non-canonical scalar field. Usually, inflation is describe by using the slow-roll parameters. The first slow-roll parameter indicates the rate of the Hubble parameter during a Hubble time as \cite{weinberg}
\begin{equation}\label{sr01}
 \epsilon=- {\dot{H} \over H^2},
\end{equation}
so that to have a quasi-de Sitter expansion this parameter should be much smaller than unity \cite{weinberg}. Using this equation and Eq.(\ref{friedmann}), one could obtain the time derivative of the scalar field in terms of the Hubble parameter and the slow-roll parameters $\epsilon$ as \cite{Unnikrishnan}
\begin{equation}\label{phidot}
\dot\phi^{2\alpha} = {2^\alpha \lambda \over \alpha} \; \epsilon \; H^2.
\end{equation}
Assuming the Hubble parameter as a function of scalar field, $H:=H(\phi)$, one has $\dot{H}=\dot\phi H_{,\phi}$. Then, the slow-roll parameter $\epsilon$ is read as
\begin{equation}\label{epsilon}
\epsilon = \left( {2^\alpha \lambda \over \alpha} \right)^{1 \over 2\alpha - 1} \;
{H_{,\phi}^{2\alpha \over 2\alpha - 1} \over H^2}
\end{equation}
The second slow-roll parameter is defined as $\eta= \ddot{\phi} / \dot{\phi}H$, where in constant-roll inflation it is assumed as a constant $\eta = \beta$. Using this definition and also Eq.(\ref{phidot}) and (\ref{epsilon}), we arrive at the following differential equation for the Hubble parameter
\begin{equation}\label{hubblede}
  H_{,\phi}^{2-2\alpha \over 2\alpha - 1} H_{,\phi\phi} = C_0 \; H,
       \qquad C_0 \equiv (2\alpha-1) \; \beta \left( {\alpha \over 2^\alpha \lambda } \right)^{1 \over 2\alpha - 1},
\end{equation}
which comes to the same differential equation for the Hubble parameter as in \cite{Motohashi}, where the authors consider the constant-roll inflation for the canonical scalar field. Since $H_{,\phi} = dH/d\phi$ there is $d\phi=dH / H_{,\phi}$. Therefore, for the second derivative of the Hubble parameter in terms of the scalar field we have $H_{,\phi\phi}= dH_{,\phi} / d\phi = H_{,\phi}dH_{,\phi} / dH$. Substituting this into the above differential equation, one arrives at
\begin{equation}\label{Hphi}
H_{,\phi}^{2\alpha \over 2\alpha -1} = {\alpha C_0 \over 2\alpha -1 } \; H^2 + C_1;
\end{equation}
where $C_1$ is the constant of integration. Taking another integration from Eq.(\ref{Hphi}), the scalar field could be expressed in terms of the Hubble parameter as
\begin{equation}\label{phiH}
\phi = \phi_0 + {H \over C_1} \;
       {}_2F_1\Big[{1 \over 2},1-{1 \over 2\alpha},{3 \over 2},{C_0 \alpha \over C_1(1-2\alpha)}\;H^2 \Big],
\end{equation}
in which $\phi_0$ is constant of integration too. From Eqs.(\ref{Hphi}) and (\ref{phiH}), it seems that every quantity could be expressed in terms of the Hubble parameter, such as the slow-roll parameter
\begin{equation}\label{epsilonH}
\epsilon = \left( {2^\alpha \lambda \over \alpha} \right)^{1 \over 2\alpha - 1} \;
         { \left( {\alpha C_0 \over 2\alpha -1 } \; H^2 + C_1 \right) \over H^2 },
\end{equation}
and also for the number of e-folds there is
\begin{eqnarray}\label{efold}
N & = & \int_{t_i}^{t_e} H dt = \int_{H_i}^{H_e} {H \over \dot{H}} dH = \int_{H_i}^{H_e} {-1 \over \epsilon \; H} \; dH \nonumber \\
     & = & -\left( \alpha \over 2^\alpha \lambda \right)^{1 \over 2\alpha-1}
           \int_{H_i}^{H_e} {H \; dH \over {\alpha C_0 \over 2\alpha -1} \; H^2 + C_1}
\end{eqnarray}
and by taking integrate one arrives at
\begin{equation}\label{efoldN}
N = -\left( \alpha \over 2^\alpha \lambda \right)^{1 \over 2\alpha-1} \; {2\alpha-1 \over 2\alpha C_0} \;
\ln\left[ {\alpha C_0 \over 2\alpha -1} \; H^2 + C_1 \right]\Bigg|_{H_i}^{H_e}
\end{equation}
where $H_e$ is the Hubble parameter at the end of inflation, and $H_i$ is the Hubble parameter at the horizon exit time. From Eq.(\ref{potential}) the potential of the scalar field could be derived only as a function of the Hubble parameter too
\begin{equation}\label{potH}
  V = 3 M_p^2 H^2 \left[1 - {(2\alpha-1) \over 3\alpha} \;
  \left( {2^\alpha \lambda \over \alpha} \right)^{1 \over 2\alpha - 1}
  { \left( {\alpha C_0 \over 2\alpha -1 } \; H^2 + C_1 \right) \over H^2 } \right]
\end{equation}

\section{Perturbations of non-canonical scalar field}
Assume a small inhomogeneity of scalar field as $\phi(t,\mathbf{x}) = \phi_0(t) + \delta\phi(t,\mathbf{x})$. This perturbation with respect to the background $\phi_0(t)$ (which from now on we omit the subscript "0") induce a perturbation to the metric because from the field equation it is clear that geometry and matter are tightly coupled together. The metric in longitudinal gauge is written as
\begin{equation}
ds^2 = \big(1 + 2\Phi(t,\mathbf{x}) \big) dt^2 - a^2(t) \big(1 - 2\Psi(t,\mathbf{x})\big)\delta_{ij} dx^i dx^j
\end{equation}
where by assuming a diagonal tensor for the spatial part of energy-momentum tensor (i.e. $\delta T^i_j \propto \delta^i_j$) there is $\Psi(t,\mathbf{x})=\Phi(t,\mathbf{x})$. \\
The action for linear scalar perturbation is derived as \cite{Garriga,Mukhanov}
\begin{equation}\label{vaction}
  S = {1 \over 2} \int  \left( v'^2 + c_s^2 v (\nabla v)^2 + {z'' \over z} v \right) d\tau d^3\mathbf{x}.
\end{equation}
in which $v \equiv z \zeta$, and $\zeta$ is the curvature perturbation given by $\zeta = \Phi + H \; {\delta\phi \over \dot\phi}$, and prime denotes derivative with respect to the conformal time $\tau$, $ad\tau=dt$. Also, $c_s$ is the sound speed of the model which is stated as
\begin{equation}\label{cs}
  c_s = {1 \over \sqrt{2\alpha -1}}
\end{equation}
which is a constant. To get a physical sound speed, the constant $\alpha$ should be always positive and bigger than $0.5$, and also to not exceed the speed of light it should be bigger than one. \\
The quantity $z$ is known as the Mukhanov variable that for our model it is defined as \cite{Garriga,Mukhanov}
\begin{equation}
  z^2 \equiv {2\alpha a X \over c_s^2 H^2 } \; \left( {X \over M^4} \right)^{\alpha-1}.
\end{equation}
From Eq.(\ref{vaction}), the equation for perturbation quantity $v$ becomes
\begin{equation}\label{vequation}
{d^2 \over d\tau^2} v(\tau,\mathbf{x}) - c_s^2 \nabla^2 v(\tau,\mathbf{x}) - {z'' \over z} \; v(\tau,\mathbf{x})=0.
\end{equation}
and by utilizing the Fourier mode, one arrives at
\begin{equation}\label{vkequation}
{d^2 \over d\tau^2} v_k(\tau) +\left( c_s^2 k^2 - {1 \over z}\;{d^2z \over d\tau^2} \right) \; v_k(\tau)=0.
\end{equation}
Before discussing the solution of the above equation, we try to compute the term $z''/z$. From the definition of $z$ and the slow-roll parameters of the previous section, there is
\begin{equation}
  z \equiv \sqrt{2\alpha \over 2^\alpha \lambda} \; {a \dot\phi^2 \over c_s H}
\end{equation}
\begin{equation}
  {dz \over d\tau} = z \big( aH \big) \Big[ 1 + \alpha \eta + \epsilon \Big]
\end{equation}
To calculate the second derivative, we first need to obtain the derivative of the slow-roll parameters with respect to the conformal time
\begin{equation}
  {d\epsilon \over d\tau} = \big( aH \big) \Big[ 2\alpha \eta\epsilon + \epsilon^2 \Big],
\end{equation}
Then, the second derivative of the quantity $z$ with respect to the conformal time is read by
\begin{equation}
  {1 \over z}{d^2z \over d\tau^2} = \big( aH \big)^2 \Big[ 2 + 2\epsilon + 3\eta + (\alpha+2\alpha\theta) \eta \epsilon + \alpha^2 \eta^2 \Big].
\end{equation}
Here $\theta = \pm 1$, and appears through the time derivative of the scalar field (From Eqs.(\ref{phidot}) and (\ref{epsilon}), $\dot\phi^2$ are derived in terms of the scalar field. On the other hand, time derivative of $\epsilon$ could be expressed as $\dot{\epsilon} = \dot\phi \epsilon_{,\phi}$. Then, we need $\dot\phi$ which in general is $\dot\phi = \theta \; \sqrt{\mathcal{O}}$. ) \\
At subhorizon scale, where $c_sk \gg aH$ (i.e. $c_sk \gg z''/z$), the quantity $v_k$ is obtained as
\begin{equation}\label{subv}
  {d^2 \over d\tau^2} v_k(\tau) + c_s^2 k^2  v_k(\tau)=0, \quad \Rightarrow \quad
             v_k(\tau)={1 \over \sqrt{2c_s k}} e^{ic_s k \tau}.
\end{equation}
In order to find the general solution of equation (\ref{vkequation}), we make the variable changes $v_k=\sqrt{-\tau} f_k$ and $x\equiv -c_sk\tau$. After some manipulation, the find the following Bessel's differential equation
\begin{equation}\label{bessel}
  {d^2 f_k \over dx^2} + {1 \over x}{d f_k \over dx} + \Big( 1 - {\nu^2 \over x^2} \Big)f_k=0, \qquad
             {z'' \over z} = {\nu^2 - {1 \over 4} \over \tau^2 }
\end{equation}
where we use $a^2H^2 = (1+\epsilon)^2 / \tau^2$, and keeping only first order of the slow-roll parameter $\epsilon$, the parameter $\nu$ is acquired as
\begin{equation}\label{nu}
\nu^2 = {9 \over 4} + 6\epsilon + 3\alpha \; \eta + (7 \alpha + 2\alpha\theta) \; \epsilon \; \eta + (1+2\epsilon)\; \alpha^2 \eta^2.
\end{equation}
The general solution for the above differential equation is the first and second kind of Henkel function. However to have a constancy with the solution in subhorizon scale, the second kind of Henkel function should be ignored, therefore the final solution is acquired as
\begin{equation}\label{vksolution}
v_k(\tau) = {\sqrt{\pi} \over 2}\; e^{i{\pi \over 2}(\nu + {1 \over 2})} \sqrt{-\tau} \; H_\nu^{(1)}(-c_sk\tau).
\end{equation}
The spectrum of curvature perturbation is defined as
\begin{equation}
  \mathcal{P}_s = {k^3 \over 2\pi^2}\; \left| \zeta \right|^2 = {k^3 \over 2\pi^2}\; \left| {v_k \over z} \right|^2.
\end{equation}
On superhorizon scale, the asymptotic behavior of Henkel function is
\begin{equation}
  \lim_{-k\tau \rightarrow \infty}H_\nu^{(1,2)}(x) = \sqrt{{2 \over \pi}} \; {1 \over \sqrt{2c_s k}} e^{\mp (ic_s k\tau + \delta)},
         \qquad  \delta={1 \over 2} \big( \nu + {1 \over 2} \big).
\end{equation}
Then, the spectrum of curvature perturbation on superhorizon scale will be
\begin{equation}\label{psspectrum}
  \mathcal{P}_s = \left( 2^{\nu-{3 \over 2}} \Gamma(\nu) \over \Gamma(3/2) \right)^2 \; \left( H^2 \over 2\pi \sqrt{c_s(\rho+p)} \right)^2 \;  \left( c_s k \over a H \right)^{3-2\nu}
\end{equation}

\section{Consistency with observation}
An advantage of the inflationary scenario is the prediction of the quantum perturbations including scalar, vector and tensor ones. Amongst them the scalar perturbations are considered as the primary seeds for large structure formation and the tensor perturbation is interpreted as the source for the gravitational waves. During inflation these perturbations are stretched out the horizon and remain invariant. On the other hand the observational data almost indicates an  scale invariant spectrum for curvature perturbations. This scale invariant feature is described by scalar spectral index, so that it is defined as $\mathcal{P}_s = A_s^2 \left( c_s k \over a H \right)^{n_s-1}$, where $A_s^2$ is the amplitude of scalar perturbation at horizon crossing $c_sk = aH$. For $n_s=1$ the amplitude of scalar perturbation is exactly scale invariant, however the latest observational data implies that $\ln\left( A_s \times 10^{10} \right) = 3.044 \pm 0.014 $ and $n_s=0.9649 \pm 0.0042$ expressing an almost scale invariant perturbations \cite{planck}. To measure the tensor perturbations the usual procedure  is indirectly through the parameter $r$ known as the tensor-to-scalar ratio, $r=\mathcal{P}_t/\mathcal{P}_s$. The data originated by Planck-2018 show there is only an upper bound on this parameter, $r < 0.064$, and it is still not measured exactly \cite{planck}. \\
Consistency of the presented model with observational data is the main goal of this section. At the end of inflation the first slow-roll parameter reaches unity\footnote{The condition $\epsilon=1$ could be interpreted as the end of inflation like the standard inflationary scenario, or it could express the most initial time of the beginning of inflation like intermediate inflation.} in which it happens for $H=H_e$
\begin{equation}\label{epsilon=1}
\epsilon(H_e)= \left( {2^\alpha \lambda \over \alpha} \right)^{1 \over 2\alpha - 1} \;
         { \left( {\alpha C_0 \over 2\alpha -1 } \; H^2 + C_1 \right) \over H^2 }=1,
\end{equation}
which leads to the following value for the Hubble parameter at the end of inflation
\begin{equation}\label{He}
H_e^2 = \left( {2^\alpha \lambda \over \alpha} \right)^{1 \over 2\alpha-1} \; {-C_1 \over \alpha\eta - 1}.
\end{equation}
To guarantee the positiveness of $H_e^2$, the term $-C_1 \lambda / (\alpha\eta -1)$ should always be positive. At the horizon exit the slow-roll parameter $\epsilon$ is smaller than unity, and the corresponding Hubble parameter could be determine through the number of e-fold equation (\ref{efold})
\begin{equation}\label{hini}
  H^2_\star(N) = \left( {2^\alpha \lambda \over \alpha} \right)^{1 \over 2\alpha-1} \; {-C_1 \over \alpha\eta - 1} \;
  \left( {e^{2\alpha \eta N} + \alpha \eta - 1 \over \alpha \eta} \right)
\end{equation}
which is expressed in term of the number of e-fold parameter $N$. The right hand side should always be positive, so beside above restriction, the term on the parenthesis should be always positive. Note that, in order to overcome the horizon and flatness problems we need about $60-70$ number of e-fold \cite{Baumann}. \\
From Eqs.(\ref{friedmann}), (\ref{energypressure}) and (\ref{phidot}) that the time derivatives of the Hubble parameter is negative, so by passing time and approaching to the end of inflation, the Hubble parameter decreases. On the other hand, since the $\epsilon=1$ is taken as the end of inflation, it is expected that the first slow-roll parameter to be smaller than one for bigger values of the Hubble parameter, namely the horizon crossing occurs for bigger values of the Hubble parameter. This point is illustrated in Fig.\ref{epsilonH}, where by passing time and decreasing the Hubble parameter the parameter $\epsilon$ approaches one, indicating the end of inflation (the selected values for the constant in the figure is based on the results that will be determined in next lines). \\
\begin{figure}[h]
  \centering
  \includegraphics[width=10cm]{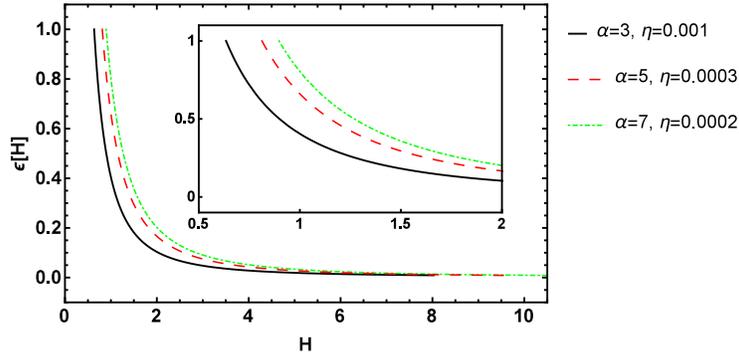}
  \caption{The figures shows the behavior of the first slow-roll parameter $\epsilon$ in terms of the Hubble parameter for different values of $\alpha$ and $\eta$.}\label{epsilonH}
\end{figure}

\begin{figure}[tbp]
\centering 
\subfigure[]{\includegraphics[width=.48\textwidth]{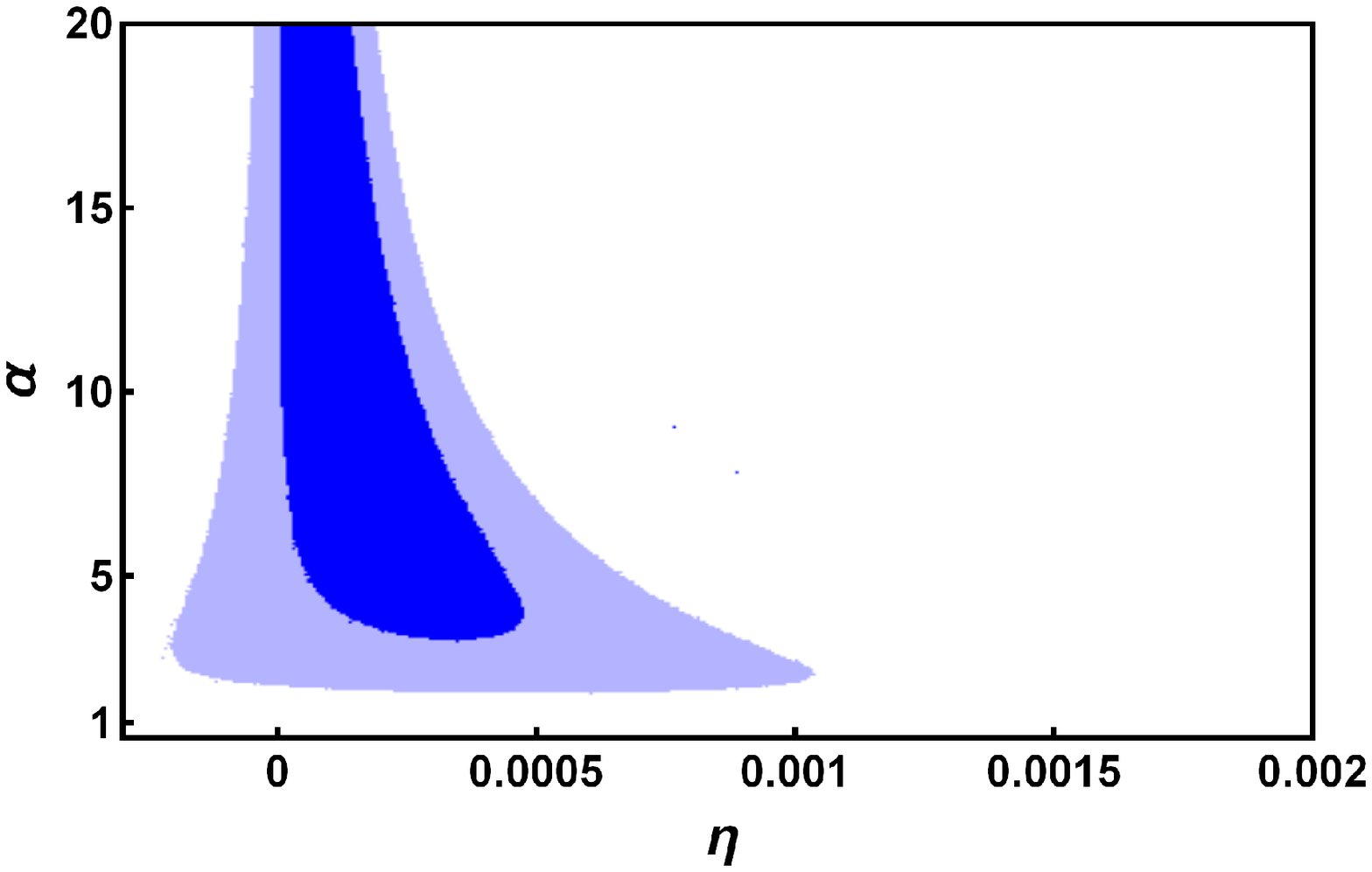}}
\subfigure[]{\includegraphics[width=.48\textwidth]{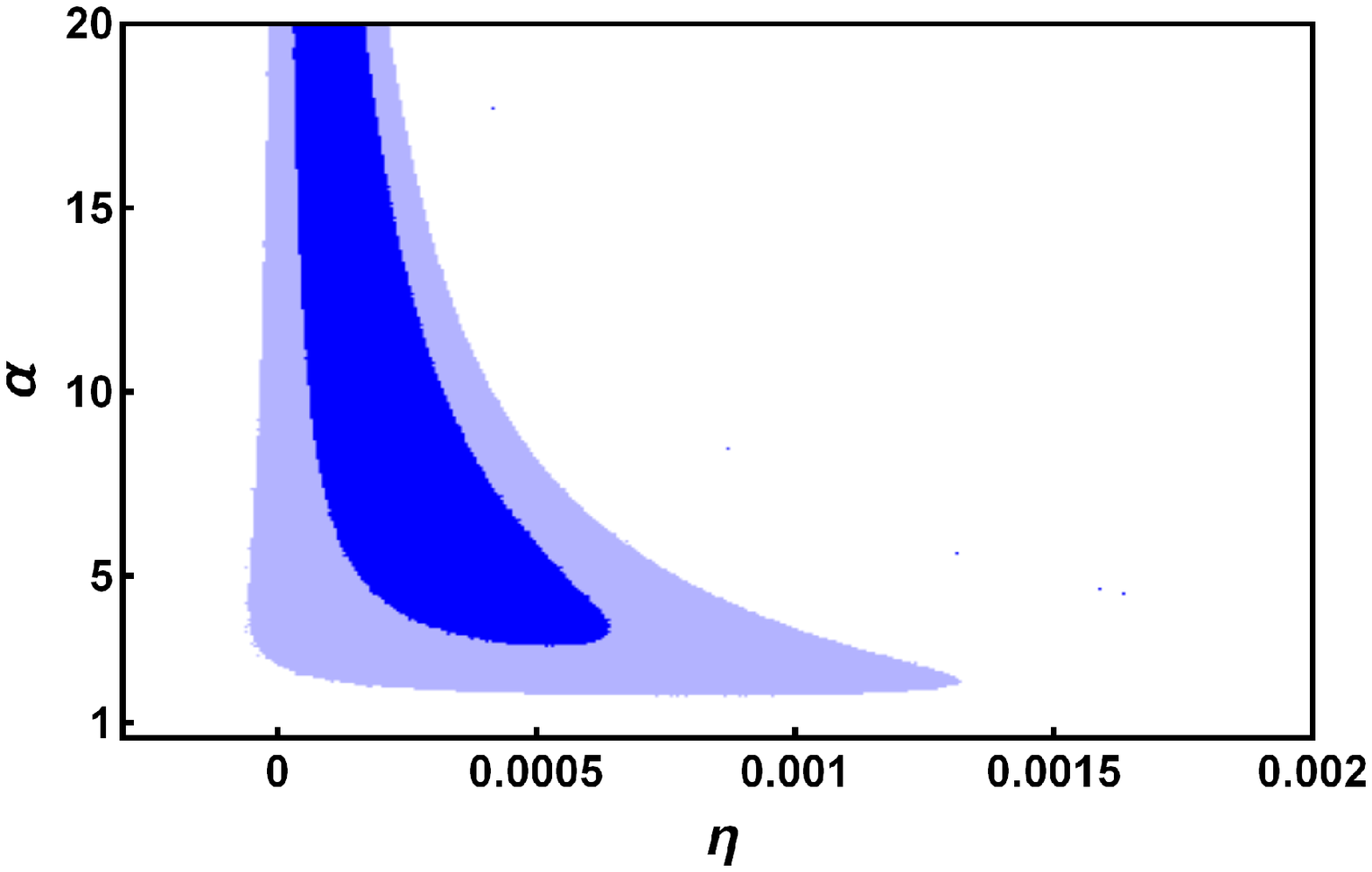}}
  \caption{The figure shows a parametric plot for the constant $\eta$ and $\alpha$ so that for these values of the constants the scalar spectral index and tensor-to-scalar ratio stands in the observational range. The figure has been plotted for $\theta=-1$ and number of e-fold: a) $N=65$, b) $N=70$. }\label{aetaplot}
\end{figure}

Using Eq.(\ref{hini}), the main perturbations parameters such as the scalar spectral index, amplitude of scalar perturbation, and tensor-to-scalar ratio could be obtained in terms of $N$ as
\begin{eqnarray}
  \epsilon(N) &=&  {\alpha \beta e^{2\alpha \beta N}  \over e^{2\alpha \beta N} + \alpha \beta - 1}, \\
  n_s(N) &=& 4 - 2\nu(N) \\
  \nu^2(N) &=& {9 \over 4} + 3 \alpha \beta + \alpha^2  \beta^2 + \Big( 6 + 7 \alpha + 2\alpha \theta + 2 \alpha^2  \beta^2\Big) \; \epsilon(N)  \\
  \mathcal{P}_s(N) &=& {1 \over 8\pi^2} \; \left( 2^{\nu(N)-{3 \over 2}} \Gamma\big(\nu(N)\big) \over \Gamma(3/2) \right)^2 \; {H^2(N) \over c_s \epsilon(N) } \\
  r(N) &=& 16 \; \left( \Gamma(3/2) \over 2^{\nu(N)-{3 \over 2}} \Gamma\big(\nu(N)\big) \right)^2 c_s \epsilon(N)
\end{eqnarray}
It is clear that the scalar spectral index and the tensor-to-scalar ratio only depend on the constant $\alpha$ and $\eta$. Using the Planck $r-n_s$ diagram, we could obtained a range of these constants that put the model predictions about $n_s$ and $r$ in the range of data. Fig.\ref{aetaplot} shows this area in which the light blue area illustrated the range related to the $95\%$ CL and the dark blue color determines values of $\alpha$ and $\eta$ that put $n_s$ and $r$ in $68\%$ CL. It shows that the second slow-roll parameter should be positive and there is no consistent result for negative $\eta$.
The second slow-roll parameter could be of order $10^{-3}$, but the best results occurs for smaller values. By increasing the constant $\eta$ the range of $\alpha$ deceases. \\
The other constants $M$ and $C_1$ appear in the amplitude of curvature perturbation and also in the potential of the scalar field through the Hubble parameter. Imposing the observational data for the amplitude of curvature perturbations determines that there should be
\begin{equation}\label{MC1}
{-C_1 M^{4(\alpha-1) \over 2\alpha-1} \over \alpha \beta -1} = \left( \alpha \over 2^\alpha \right)^{1 \over 2\alpha -1} \;
  {\mathcal{P}_s^\star \over 8\pi^2 c_s \epsilon^\star} \;
  \left( \Gamma(3/2) \over 2^{\nu^\star - {3 \over 2}} \Gamma(\nu^\star) \right)^2 \;
  {\alpha \eta \over e^{2\alpha \eta N} + \alpha \eta -1}
\end{equation}
The amplitude of the curvature perturbation is clear from data, so by choosing some values for the constants $\alpha$ and $\eta$ from Fig.\ref{aetaplot}, the right hand side would be clear. Table.\ref{MC1table} expresses this terms for different choices of $\alpha$ and $\eta$. \\

Also, the term on the left hand side of Eq.(\ref{MC1}) appears in the potential of the scalar field, in which at the time of horizon crossing, one has
\begin{equation}\label{MC1Vstar}
V^\star = 3 \left( 2^\alpha \over \alpha \right)^{1 \over 2\alpha-1} \; {-C_1 M^{4(\alpha-1) \over 2\alpha-1} \over \alpha \beta -1} \; {e^{2\alpha \eta N} \over \epsilon^\star} \;
\left[ 1 - {(2\alpha-1) \over 3\alpha} \; \epsilon^\star \right]
\end{equation}
Then, the determined values of $\alpha$ and $\eta$, the inflation energy scale is determined, that has been implied in Table.\ref{MC1table}. It is clearly seen that the energy scale of inflation is about $10^{-2}$.
\begin{table}[h]
  \centering
  \begin{tabular}{p{0.8cm}p{1.5cm}p{2.8cm}p{2.2cm}}
    \hline
    $\alpha$ & $\quad \eta$ & $\ \ -C_1 M^{4(\alpha-1) \over 2\alpha-1}$ & \qquad $V^\star$ \\
    \hline
    $3$ & $0.0002$ & $-4.56 \times 10^{-11}$ & $2.26 \times 10^{-8}$ \\
    $3$ & $0.001$  & $-3.30 \times 10^{-11}$ & $1.92 \times 10^{-8}$ \\
    $5$ & $0.0003$ & $-5.73 \times 10^{-11}$ & $2.94 \times 10^{-8}$ \\
    $5$ & $0.0005$ & $-4.69 \times 10^{-11}$ & $2.66 \times 10^{-8}$ \\
    $7$ & $0.0002$ & $-6.73 \times 10^{-11}$ & $3.53 \times 10^{-8}$ \\
    $7$ & $0.0004$ & $-5.58 \times 10^{-11}$ & $3.21 \times 10^{-8}$ \\
    \hline
  \end{tabular}
  \caption{Determining the constant term $C_1 M^{4(\alpha-1) \over 2\alpha-1}$ by using Fig.\ref{aetaplot}, and data for the curvature perturbations. Then, the energy scale of inflation is illustrated foe these values of the constants. The result have been derived for number of e-fold $N=65$.}\label{MC1table}
\end{table}

\section{Attractor Behavior}
\begin{figure}[h]
  \centering
  \includegraphics[width=10cm]{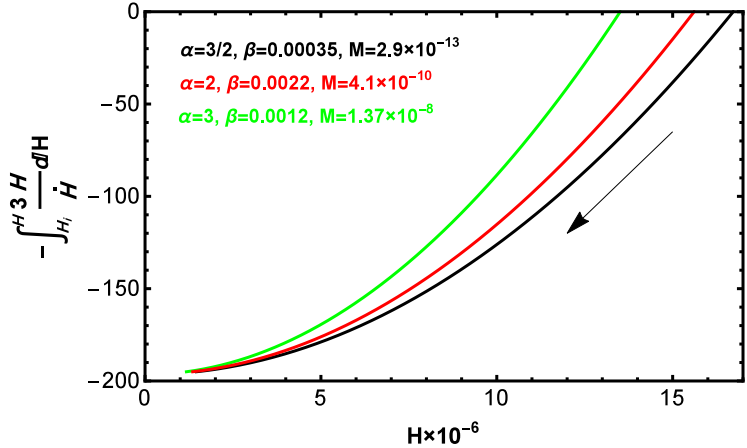}
  \caption{The power of the exponential term in Eq.(\ref{attractor-eq}) during the inflation. By passing time and approaching to the end of inflation, the power increases with negative sign. Then, the perturbation $\delta H$ exponentially decreases during inflation and the solution is attractive. }\label{attractor}
\end{figure}

Considering the attractor behavior of the solution will be performed following the same process as mentioned in \citep{lyth,Odintsov-a,Odintsov-b} which is a simple approach and also effective as one works in Hamilton-Jacobi formalism. In this formalism it is assumed that there is a homogenous perturbation to the Hubble parameter as $H=H_0+\delta H$. Now by introducing it into the Hamilton-Jacobi equation (\ref{potential}), and keeping the terms up to the first order of $\delta H$, one arrives at
\begin{equation}
  {\delta H ' \over \delta H} = 3 \left( {2^\alpha \lambda \over \alpha} \right) \; {H_0 \over \dot\phi_0^{2\alpha}} \; H'_0,
\end{equation}
By taking integrate from the above equation, we have
\begin{equation}\label{attractor-eq}
  \delta H = \delta H_i \exp\left( - 3 \int_{H_i}^{H} {H \over \dot{H}} \; dH \right).
\end{equation}
where $\delta H_i$ is the perturbation at initial time. From above equation, it is realized that by approaching to the end of inflation the term within the power of the exponential term grows, and the negative sign brings out this fact that the perturbation $\delta H$ decreases continuously. This feature is determined in Fig.\ref{attractor}, where the integral has been plotted for different values of $\alpha$, $\beta$ and $M$ and for all cases the term deceases during the inflation. Therefore, it is resulted that the model satisfies the attractor behavior.

\section{Conclusion}
This work was relied  on a scenario of constant-roll evolution of the initial Universe. It has been supposed that inflation has derived considering a modification in the kinetic term of the Lagrangian called the non-canonical model of the scalar field. In the constant-roll scenario it has assumed that, the second slow-roll parameter, i.e. $\eta$, has to be a constant and therefore we had to recalculate necessary inflationary parameters. By calculating the slow-roll parameters in term of the Hubble parameter we could derive a differential equation for the Hubble parameter which could lead to the corresponding differential equation for canonical scalar field model by taking $\alpha=1$. On the other side, by taking into account the assumption of constant-roll method the perturbation equation should be recalculate again. Doing so obviously there are some modifications in the amplitude of scalar perturbation and scalar spectral index lead to the appearance of the second order of $\eta$ in our investigation. \\

By finding an exact solution, which is one of the main advantages and triumphs of this approach, it was realized that every parameter of the model can be expressed in terms of the Hubble parameter. Therefore, comparing our approach with the usual Hamilton-Jacobi formalism of inflation the Hubble parameter here, behaves as same as that scalar field in that approach. The importance of this result goes back to this fact that there is no need to introducing an ansatz for Hubble parameter based on the scalar field. The condition $\epsilon = 1$ clarified the Hubble parameter at the end of inflation, i.e. $H_e$, and using the number of e-fold the Hubble parameter has determined in terms of the number of e-fold and other constant parameters of the model. Utilizing this result, we have expressed the main perturbation parameters of the model in terms of the number of e-fold to obtain better estimations comparing to observationas. \\

It has been shown that the scalar spectral index and the tensor-to-scalar ratio at the  horizon exit can be obtained in terms of two free parameters of the model namely $\alpha$ and $\eta$. Then, by making use of the $r-n_s$ diagram originated from the Planck 2018, the proper ranges based on these two parameters have been estimated and depicted in Fig.\ref{aetaplot}. It has been noticed for any values of $\alpha$ and $\eta$ in the best estimated range both the scalar spectral index and tensor-to-scalar  ration are in good agreement with observational data. The results show that the second slow-roll parameter $\eta$ should be constant, as it should,  and is of order $10^{-4}$, and by increasing the constant $\alpha$ the range of $\eta$ becomes smaller. On the other hand, applying the data for the amplitude of curvature perturbation, the constant $C_1 M^{4(\alpha-1) \over 2\alpha-1}$ has determined, and imposing this result in the potential of the scalar field it has found out that the energy scale of inflation is around $V^{\star {1/4}} \propto 10^{-2}$.\\

Ultimately, the attractor behavior of the model by assuming a homogenous perturbation to the Hubble parameter has been investigated. By introducing into the Hamilto-Jacobi equation, it leads to a differential equation of the homogenous perturbation which from Fig.\ref{attractor} it was obvious that the perturbation decreases exponentially by approaching to the end of inflation, that indicates the  attractor behavior of  solution.

\begin{acknowledgments}
AM would like to thanks the "Ministry of Science, Research and Technology" of Iran for financial support, and Prof. R. Casadio for hospitality during the visit.
HS thanks A. Starobinsky for very constructive discussions about inflation during Helmholtz International Summer School  2019 in Russia. He grateful G. Ellis, A. Weltman, and UCT for arranging his short visit, and for enlightening discussions about cosmological fluctuations and perturbations for both large and local scales.
\end{acknowledgments}



\begin{thebibliography}{00}




\bibitem{Guth} A. H. Guth, \emph{Inflationary universe: A possible solution to the horizon and flatness problems}, Phys. Rev. D 23 (1981) 341.

\bibitem{Linde} A. D. Linde, \emph{A new inflationary universe scenario: a possible solution of the horizon, flatness, homogeneity, isotropy and primordial monopole problems}, Phys. Lett. B 108, 389 (1982).

\bibitem{Albrecht} A. Albrecht and P. J. Steinhardt, \emph{Cosmology for Grand Unified Theories with Radiatively Induced Symmetry Breaking}, Phys. Rev. Lett. 48, 1220 (1982).

\bibitem{Lindea} A. D. Linde, \emph{Chaotic Ination}, Phys. Lett. B 129, 177 (1983).

\bibitem{Lindeb} A. D. Linde,  \emph{Particle Physics and Inflationary Cosmology}, Harward Academic, Chur, Switzer-
land (1990).

\bibitem{Armendariz-Picon} C. Armendariz-Picon, T. Damour and V. F. Mukhanov, \emph{k-inflation}, Phys. Lett. B 458, 209 (1999)

\bibitem{Garriga} Jaume Garriga, V.F. Mukhanov, \emph{Perturbations in k-inflation}, Phys. Lett. B 458, 219 (1999).

\bibitem{Maartens} R. Maartens, D. Wands, B. A. Bassett, and I. P. C. Heard, \emph{Chaotic inflation on the brane}, Phys. Rev. D 62, 041301 (2000).

\bibitem{Golanbari} T. Golanbari, A. Mohammadi, Kh. Saaidi, \emph{Brane inflation driven by noncanonical scalar field}, Phys. Rev. D 89, 103529 (2014).

\bibitem{Abolhasani} A. A. Abolhasani, R. Emami, H. Firouzjahi, \emph{Primordial anisotropies in gauged hybrid inflation}, JCAP 05, 016 (2014).

\bibitem{Maeda} Kei-ichi Maeda, Kei Yamamoto, \emph{Stability analysis of inflation with an SU (2) gauge field}, JCAP 12, 018 (2013).

\bibitem{Alexander} S. Alexander, D. Jyoti, A. Kosowsky, A. Marciano, \emph{Dynamics of gauge field inflation}, JCAP 05, 005 (2015).

\bibitem{Tirandari} M. Tirandari, Kh. Saaidi, \emph{Anisotropic inflation in Brans–Dicke gravity}, Nucl. Phys. B 925, 403 (2017).

\bibitem{Berera} A. Berera, \emph{Warm inflation}, Phys. Rev. Lett. 75, 3218 (1995).

\bibitem{Bereraa} A. Berera, \emph{Warm inflation in the adiabatic regime-a model, an existence proof for inflationary dynamics in quantum field theory}, Nucl. Phys. 585, 666 (2000).

\bibitem{Taylor} A. N. Taylor and A. Berera, \emph{Perturbation spectra in the warm inflationary scenario}, Phys. Rev. D 62, 083517 (2000).

\bibitem{Hall} L. M. H. Hall, I. G. Moss, and A. Berera, \emph{Scalar perturbation spectra from warm inflation}, Phys. Rev. D 69, 083525 (2004).

\bibitem{Bastero} M. Bastero-Gil and A. Berera, \emph{Determining the regimes of cold and warm inflation in the supersymmetric hybrid model}, Phys. Rev. D 71, 063515 (2005).

\bibitem{Sayar} K. Sayar, A. Mohammadi, L. Akhtari, Kh. Saaidi, \emph{Hamilton-Jacobi formalism to warm inflationary scenario}, Phys. Rev. D 95, 023501 (2017).

\bibitem{Akhtari} L. Akhtari, A. Mohammadi, K. Sayar, Kh. Saaidi, \emph{Hamilton-Jacobi formalism to warm inflationary scenario}, Astropart. Phys. 90, 28 (2017).

\bibitem{Chen:2009we}
  X.~Chen and Y.~Wang,
\emph{Large non-Gaussianities with Intermediate Shapes from Quasi-Single Field Inflation},
  Phys.\ Rev.\ D {\bf 81}, 063511 (2010).

  \bibitem{Chen:2017ryl}
  X.~Chen, Y.~Wang and Z.~Z.~Xianyu,
\emph{Schwinger-Keldysh Diagrammatics for Primordial Perturbations,}
  JCAP {\bf 1712}  no.12,  006 (2017).

\bibitem{Golovnev:2008cf}
A.~Golovnev, V.~Mukhanov, and V.~Vanchurin, {\it {Vector Inflation}},  {
  JCAP} {\bf 0806}  009, (2008).

\bibitem{Adshead:2012kp}
P.~Adshead and M.~Wyman, {\it {Chromo-Natural Inflation: Natural inflation on a
  steep potential with classical non-Abelian gauge fields}},  {
  Phys.Rev.Lett.} {\bf 108}  261302 (2012).

\bibitem{weinberg} S. Weinberg, \emph{Cosmology}, Oxford University Press, USA (2008).

\bibitem{PhysRevLett112011302}
David I. Kaiser and Evangelos I. Sfakianakis,
``Multifield Inflation after Planck: The Case for Nonminimal Couplings,''
 Phys. Rev. Lett. \textbf{112}, 011302 (2014)

\bibitem{Emami:2013lma}
  R.~Emami,
  ``Spectroscopy of Masses and Couplings during Inflation,''
  JCAP {\bf 1404}, 031 (2014).

\bibitem{Sheikhahmadi:2019xkx} H.~Sheikhahmadi,
``Schwinger-Keldysh mechanism in extended quasi single field inflation,'' { \em Eur. Phys. J. C}  \textbf{79},  451, (2019).

\bibitem{Baumann:2011nk}
  D.~Baumann and D.~Green,
  ``Signatures of Supersymmetry from the Early Universe,''
  Phys.\ Rev.\ D {\bf 85}, 103520 (2012).

\bibitem{Chen:2012ge}
  X.~Chen and Y.~Wang,
  ``Quasi-Single Field Inflation with Large Mass,''
  JCAP {\bf 1209}, 021 (2012).


\bibitem{Sefusatti:2012ye}
  E.~Sefusatti, J.~R.~Fergusson, X.~Chen and E.~P.~S.~Shellard,
  ``Effects and Detectability of Quasi-Single Field Inflation in the Large-Scale Structure and Cosmic Microwave Background,''
  JCAP {\bf 1208}, 033 (2012).

\bibitem{Baumann} D. Baumann,  \emph{TASI Lectures on Inflation}, [arXiv:0907.5424v2].



\bibitem{1a} 
  J.~De-Santiago, J.~L.~Cervantes-Cota and D.~Wands,
  ``Cosmological phase space analysis of the F(X) - V($\phi$) scalar field and bouncing solutions,''
  Phys.\ Rev.\ D {\bf 87}, no. 2, 023502 (2013)
  doi:10.1103/PhysRevD.87.023502
  [arXiv:1204.3631 [gr-qc]].
\bibitem{2a} 
  W.~Fang, H.~Q.~Lu and Z.~G.~Huang,
  ``Cosmologies with general non-canonical scalar field,''
  Class.\ Quant.\ Grav.\  {\bf 24}, 3799 (2007)
  doi:10.1088/0264-9381/24/15/002
  [hep-th/0610188].

\bibitem{3a}
  V.~F.~Mukhanov and A.~Vikman,
  JCAP {\bf 0602}, 004 (2006)
  doi:10.1088/1475-7516/2006/02/004
  [astro-ph/0512066].

\bibitem{5a}
  K.~Rezazadeh, K.~Karami and P.~Karimi,
  ``Intermediate inflation from a non-canonical scalar field,''
  JCAP {\bf 1509}, no. 09, 053 (2015)
  doi:10.1088/1475-7516/2015/09/053
  [arXiv:1411.7302 [gr-qc]].


\bibitem{6a}
  S.~Unnikrishnan, V.~Sahni and A.~Toporensky,
  ``Refining inflation using non-canonical scalars,''
  JCAP {\bf 1208}, 018 (2012)
  doi:10.1088/1475-7516/2012/08/018
  [arXiv:1205.0786 [astro-ph.CO]].

\bibitem{7a}
  S.~Unnikrishnan and V.~Sahni,
  ``Resurrecting power law inflation in the light of Planck results,''
  JCAP {\bf 1310}, 063 (2013)
  doi:10.1088/1475-7516/2013/10/063
  [arXiv:1305.5260 [astro-ph.CO]].


\bibitem{8a}
  X.~M.~Zhang and j.~Y.~Zhu,
  Phys.\ Rev.\ D {\bf 90}, no. 12, 123519 (2014)
  doi:10.1103/PhysRevD.90.123519
  [arXiv:1402.0205 [gr-qc]].





\bibitem{Unnikrishnan} S. Unnikrishnan, V. Sahni and A. Toporensky, \emph{Refining inflation using non-canonical scalars}, JCAP 08, 018 (2012).

\bibitem{Kinney} W. H. Kinney, \emph{Horizon crossing and inflation with large $\eta$}, Phys. Rev. D 72, 023515 (2005).

\bibitem{Namjoo} M. H. Namjoo, H. Firouzjahi and M. Sasaki, \emph{Violation of non-Gaussianity consistency relation in a single-field inflationary model}, Europhys. Lett. 101, 39001 (2013).



\bibitem{Martin} J. Martin, H. Motohashi and T. Suyama, \emph{Ultra slow-roll inflation and the non-Gaussianity consistency relation}, Phys. Rev. D 87, 023514 (2013).

\bibitem{Motohashi} H. Motohashi, A. A. Starobinsky, and J. Yokoyama, \emph{Inflation with a constant rate of roll}, JCAP 1509, 018 (2015).


\bibitem{Salopek} D. S. Salopek and J. M. Stewart, \emph{Hamilton-Jacobi theory for general relativity with matter fields}, Classical Quantum Gravity 9, 1943 (1992).

\bibitem{Liddle} A. R. Liddle, P. Parsons, and J. D. Barrow, \emph{Formalizing the slow-roll approximation in inflation}, Phys. Rev. D 50, 7222 (1994).

\bibitem{Kinneya} William H. Kinney, \emph{Hamilton-Jacobi approach to non-slow-roll inflation}, Phys. Rev. D 56, 2002 (1997).

\bibitem{Guo} Zong-Kuan Guo, Yun-Song Piao, Rong-Gen Cai, and Yuan-Zhong Zhang, \emph{Inflationary Attractor from Tachyonic Matter}, Phys. Rev. D 68, 043508 (2003).

\bibitem{Aghamohammadi} A. Aghamohammadi, A. Mohammadi, T. Golanbari, and Kh. Saaidi, \emph{Hamilton-Jacobi Formalism for Tachyon Inflation}, Phys. Rev. D 90, 084028 (2014).

\bibitem{Saaidi} Kh. Saaidi, A. Mohammadi, and T. Golanbari, \emph{Light of Planck-2015 on Non-Canonical Inflation}, Adv. High Energy Phys. 2015, 1 (2015).

\bibitem{Sheikhahmadi} H. Sheikhahmadi, E. N. Saridakis, A. Aghamohammadi, and K. Saaidi, \emph{
Hamilton-Jacobi formalism for inflation with non-minimal derivative coupling}, J. Cosmol. Astropart. Phys. {\bf 10}, 021 (2016) .

\bibitem{Motohashi-b} H. Motohashi, A. A. Starobinsky, \emph{$f(R)$ constant-roll inflation}, Eur.Phys.J. C {\bf 77}, 538 (2017).

\bibitem{Nojiri} S. Nojiri, S. D. Odintsov, V. K. Oikonomou, \emph{Constant-roll Inflation in $f(R)$ Gravity}, Class. Quant. Grav. {\bf 34} (2017) 245012.

\bibitem{Odintsov-c} S. D. Odintsov, V. K. Oikonomou, L.Sebastiani, \emph{Unification of constant-roll inflation and dark energy with logarithmic $R^2$-corrected and exponential $f(R)$ gravity}, Nucl. Phys. B {\bf 923}, 608 (2017).

\bibitem{Oikonomou-a} V. K. Oikonomou, \emph{Reheating in constant-roll $f(R)$ gravity}, Mod. Phys. Lett. A {\bf 32}, 1750172 (2017).

\bibitem{karam} A. Karam, L. Marzola, T. Pappas, A. Racioppi, K. Tamvakis, \emph{Constant-Roll (Quasi-)Linear Inflation}, arXiv:1711.09861v1 [astro-ph].

\bibitem{Odintsov-a} S. D. Odintsov, V. K. Oikonomou, \emph{Inflationary Dynamics with a Smooth Slow-Roll to Constant-Roll Era Transition}, JCAP {\bf 04}, 041 (2017).

\bibitem{Odintsov-b} S. D. Odintsov, V. K. Oikonomou, \emph{Inflation with a Smooth Constant-Roll to Constant-Roll Era Transition}, Phys. Rev. D {\bf 96} (2017) 024029.

\bibitem{Oikonomou-b} V. K. Oikonomou, \emph{A Smooth Constant-Roll to a Slow-Roll Modular Inflation Transition}, Int. J. Mod. Phys. D {\bf 27} (2018) 1850009.

\bibitem{Mukhanov} V. F. Mukhanov, H. A. Feldman, R. H. Brandenberger, \emph{Theory of cosmological perturbations}, Phys. Rep. 215, 203 (1992).

\bibitem{planck} Y. Akrami, et al. (Planck Collaboration),  \emph{Planck 2018 results. X. Constraints on inflation}, arXiv:1807.06211 (2018).

\bibitem{lyth} D. H. Lyth and A. R. Liddle, \emph{ The Primordial Density Perturbation}, (Cambridge University
Press, Cambridge, England, 2009).





\end{thebibliography}



\end{document}